\documentclass[5p,times]{elsarticle}

\usepackage{lineno,hyperref}
\modulolinenumbers[5]

\journal{Journal of \LaTeX\ Templates}

\usepackage{amssymb,amsmath,amsfonts,amsbsy,graphicx}
\usepackage{color}
\usepackage{enumerate}

\newcommand{\dis}[1]{\begin{equation}\begin{split}#1\end{split}\end{equation}}
\newcommand{\be}{\begin{equation}}
\newcommand{\ee}{\end{equation}}
\def\bea{\begin{eqnarray}}
\def\eea{\end{eqnarray}}

\newcommand{\eq}[1]{Eq.~(\ref{#1})}

\newcommand{\bfrac}[2]{{\left(\frac{#1}{#2} \right)  }}
\newcommand{\VEV}[1]{\langle #1 \rangle}

\newcommand\tev{\,{\rm TeV}}
\newcommand\gev{\,{\rm GeV}}

\newcommand{\Mp}{M_{\rm P}}

\newcommand{\mpl}{m_{\rm Pl}}

\newcommand{\Treh}{T_{\rm reh}}

\usepackage{slashed} 

\begin{document}

\begin{frontmatter}

\title{Non-thermal WIMP baryogenesis}

\author[SKKU]{Ki-Young Choi}
\cortext[mycorrespondingauthor]{Corresponding author}
\ead{kiyoungchoi@skku.edu}

\author[SNUT]{Sin Kyu Kang}
\ead{skkang@snut.ac.kr}

\author[SKKU]{Jongkuk Kim\corref{mycorrespondingauthor}}
\ead{jongkukkim@skku.edu}

\address[SKKU]{Department of Physics, BK21 Physics Research Division, Institute of Basic Science, Sungkyunkwan University, Suwon 440-746, South Korea}
\address[SNUT]{School of Liberal Arts, Seoul-Tech, Seoul 139-743, Korea}

\begin{abstract}
We propose a model of  baryogensis achieved by the annihilation of non-thermally produced  WIMPs from decay of heavy particles, which can result in low reheating temerature. 
Dark matter (DM) can be produced non-thermally  during  a reheating period created by the decay of long-lived heavy particle, and subsequently re-annihilate to lighter particles even after the thermal freeze-out. The re-annihilation of DM provides the observed baryon asymmetry  as well as the correct relic density of DM. 
We investigate how wahout effects can affect the generation of the baryon asymmetry and study a model suppressing them.
In this scenario, we find that DM can be heavy enough and its annihilation cross section  can also be larger than that adopted in the usual thermal WIMP baryogenesis.

\end{abstract}


\end{frontmatter}


\section{Introduction}
\label{intro}
The baryon density at present inferred from Cosmic Microwave Background (CMB) anisotropy and Big Bang Nucleosynthesis (BBN) is~\cite{Ade:2015xua}
\dis{
\Omega_B h^2 = 0.0223\pm0.0002,
}
which corresponds to the baryon asymmetry
\dis{
Y_B\equiv \frac{n_B}{s} \simeq 0.86 \times 10^{-10}, \label{Ybaryon}
}
where $n_B$ and $s$ is the baryon number density and entropy density respectively.
There are many suggested models for baryogenesis. One of them is the thermal weakly interacting massive particle (WIMP) baryogenesis \cite{wimp-baryogenesis,McDonald:2011zza, Cui:2011ab, Cui:2012jh}, which has been paid much attention for past  few years thanks to the intriguing coincidence of  the observed baryon and dark matter (DM) abundances, $\Omega_B\simeq 5 \Omega_{DM}$.
WIMP miraculously accounts for  $\Omega_{DM}$, and may play a role in generation of baryon asymmetry.
The WIMP baryogenesis mechanism~\cite{Cui:2011ab} uses the WIMP dark matter annihilation during thermal freeze-out. Baryogenesis is successfully achieved because the WIMP annihilations violate baryon number, C and CP, and the out-of-equilibrium is attained when the DM number density is deviated from the thermal equilibrium. For this scenario to be effective,  the temperature of the Universe must be larger than the freeze-out temperature of DM which is $T_{\rm fr}\simeq m_\chi/20 $.  Therefore there is a limitation for low-reheating temperature.  

In new physics beyond the standard model (SM),
there are many long-lived massive particles (we call it $\phi$ afterwards) that can dominate the energy density of the Universe, and decay, such as inflaton, moduli, gravitino, axino, curvaton, and  etc~\cite{Baer:2014eja}.
These particles interact  very weakly  with visible sector and thus decay very late in the  Universe. The lifetime can be longer than $10^{-7} \sec$ which corresponds to the cosmic temperature around $1\,\gev$, which  is  far after the electroweak phase transition and freeze-out of  WIMP DM with mass $m_\chi \sim\mathcal{O}( \tev)$, whose freeze-out temperature is around $m_\chi/20$.
Then, in the models with such a long-lived particle, the reheating temperature can be low enough.
However, with such a low-reheating temperature, the relic abundance of DM can not be explained
in simple models for  thermal WIMP freeze-out. In addition, it is questionable whether baryon asymmetry can be successfully generated in models with low-reheating temperature.

Since the primodial asymmetry generated is diluted during the late time reheating, new generation of asymmetry is required. At the low temperature below the electorweak scale,  leptogenesis does not work since  the conversion of  lepton asymmetry to baryon asymmetry  via Shpaleron processes is effective at  temperatures above the electorweak scale. 
Thus, alternative to leptogenesis is demanded to generate baryon asymmetry in models with low-reheating temperature.
A direct generation of baryon asymmetry~\cite{Kim:2004te} may be  possible without the help of  Sphaleron processes.

The aim of this letter is to propose a possible way 
to generate baryon asymmetry applicable to models with low reheating temperature. 
We will show that DM can be produced from heavy long-lived unstable particles and then both
baryon and DM abundances can be achieved by the re-annihilation of DM.
While the SM particles produced from the decay of $\phi$ are thermalized quickly and find themselves in the thermal equilibrium, the interactions of  DM are so slow that can stay in the out-of-equilibrium state until their re-annihilation. After re-annihilation, the dark matter relic density is fixed~\cite{Fujii:2002kr,Pallis:2004yy,Seto:2005pj,Nakamura:2007wr,Choi:2008zq,Bernal:2012gv,Bernal:2013bga}.
As will be shown later, in this scenario, 
Sakharov conditions \cite{sakharov}  are satisfied with the violations of C and CP as well as B number during the re-annihilation of the non-thermal WIMP DMs~\footnote{A leptogenesis at the reheating era was considered~\cite{Hamada:2015xva,Hamada:2016npz}. Here they consider the SM particles from the inflaton decay are out-of equilibrium until the thermalization. During the scattering process, the asymmetry is generated in the SM sector.} which are out-of-equlibrium.

This letter is organized as follows.
 In Section~\ref{NTWIMP}, we show how non-thermal WIMP  can generate baryon asymmetry.
 Numerical results are presented in Section~\ref{results}.
 A simple model to successfully achieve non-thermal WIMP baryogenesis is provided in Section~\ref{model}. We discuss how washout can be suppressed before the baryon asymmetry is generated.
Conclusions are given in Section~\ref{Con}.

\section{Non-thermal WIMP Baryogenesis}
\label{NTWIMP}

\begin{figure*}[!t]
\begin{center}
\begin{tabular}{cc} 
 \includegraphics[width=0.45\textwidth]{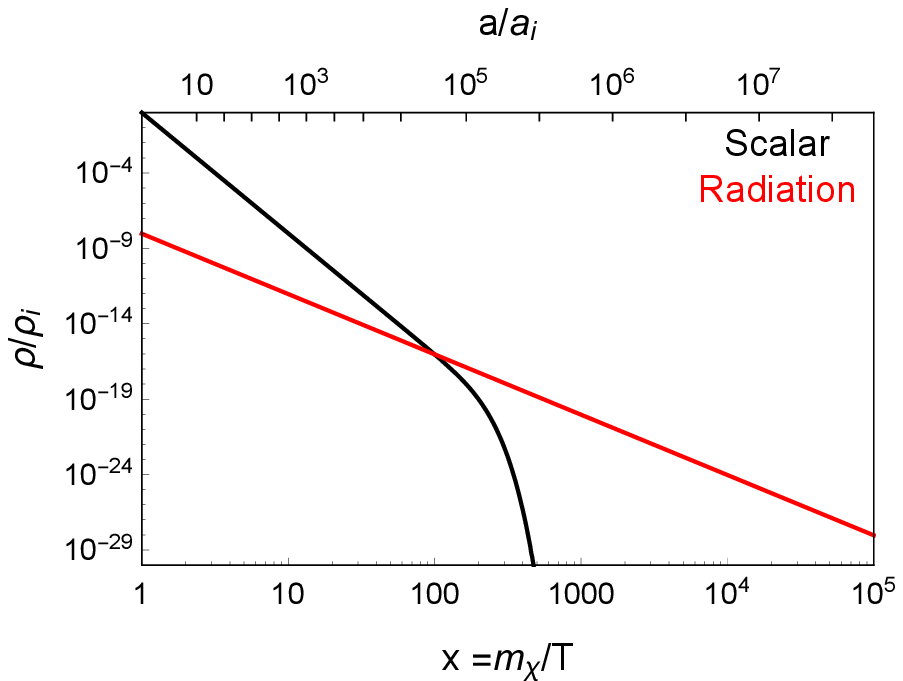}
 &
 \includegraphics[width=0.45\textwidth]{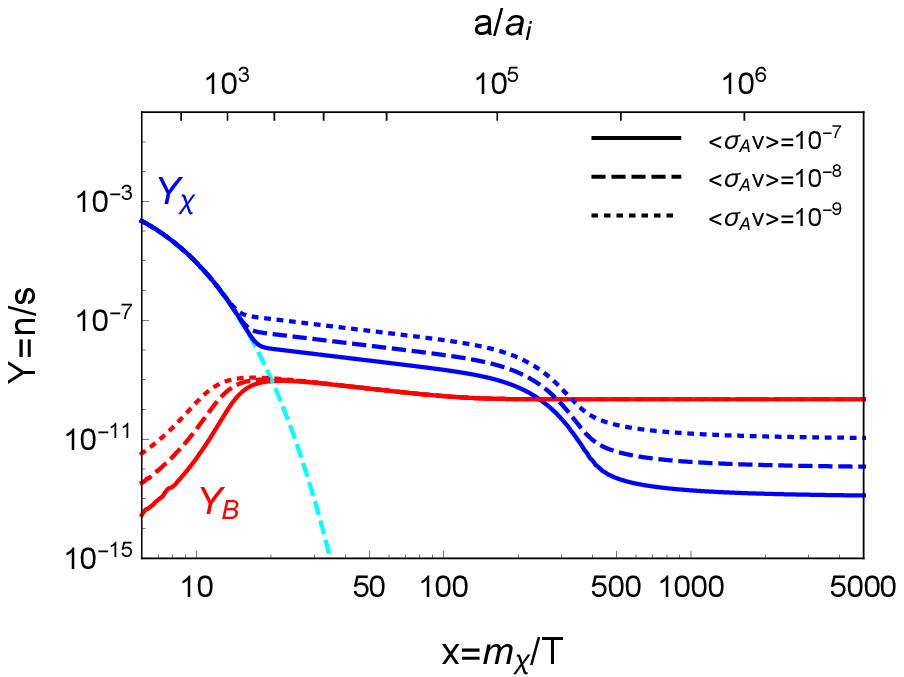}
 \end{tabular}
\end{center}
\caption{The evolution of the energy density of $\phi$  and radiation (left panel), and $Y_\chi$ and $Y_B$ (right panel) as a function of $x=m_\chi/T$ or $a/a_i$ with $m_\chi=2\tev$. As inputs we have taken  $m_\phi =5\,{\rm TeV}$, $\Treh = 20\,{\rm GeV}$ (corresponding to $x\simeq 100$ or $a/a_i\simeq 10^5$), $f_\chi=0.01$ and  $\VEV{\sigma_A v} =10^{-7}$ (soild), $10^{-8}$ (dashed), $10^{-9}$ (dotted)  ${\rm GeV^{-2}}$.  In the calculation of $Y_B$, we have used $\epsilon=0.001$, $m_\psi=3\tev$ and  $\VEV{\sigma_{\slashed{B}} v}=5\times10^{-3}\VEV{\sigma_A v}$. In the right panel, the cyan dashed line corresponds to the equlibrium number density of DM.}
\label{Fig:rho_Y}
\end{figure*}

We begin by considering a long-lived heavy particle, $\phi$, so that the corresponding reheating temperature is relatively low. Using a sudden-decay approximation, the relation between the reheating temperature and the lifetime, $\tau_\phi$, is roughly given by
\dis{
{\Treh} \simeq \bfrac{90}{\pi^2g_*}^{1/4}\sqrt{\Mp \Gamma_\phi} \simeq 2.5\gev\times  \bfrac{10^{-7}\sec}{\tau_\phi}^{1/2},
}
where we used the decay width $\Gamma_\phi = \tau_\phi^{-1}$.
For a heavy scalar particle whose interactions to SM particles are suppressed by a certain high scale $\Lambda$, its decay rate and lifetime are roughly given by
\dis{
\Gamma_\phi \sim \frac{1}{64\pi} \frac{m_\phi^3}{\Lambda^2}, \quad \textrm{and} \quad \tau_\phi \sim 10^{-7}\sec \bfrac{1\tev}{m_\phi}^3 \bfrac{\Lambda}{10^{12}\gev}^2,
}
respectively. Therefore in the following we will focus on the case of $m_\phi \simeq \mathcal{O} (\tev)$ with $\Lambda = 10^{12}\gev$,  which gives a reheating temperature lower than the WIMP  freeze-out temperature.

The decay of $\phi$ is continuous and even before reaching the lifetime, i.e. when $t\ll \tau_\phi$, the relativistic particles and DMs are produced continuously.  Right after the production, they are non-thermal with the   energy of $E\sim m_\phi/2$. The SM particles which have gauge interactions and large Yukawa couplings scatter efficiently and quickly settle  down to the thermal equilibrium with corresponding temperature $T$, defined by
\dis{
\rho_r = \frac{\pi^2}{30}g_* T^4,
}
where $\rho_r$ is the energy density of the relativistic particles in the thermal equilibrium with the effective degrees of freedom $g_*$.
However for  DMs which have weak interactions, their scatterings are relatively slow and do not lead to the thermal equilibrium quickly.  Instead they stay in the out-of-equilibrium until the re-annihilation happens efficiently.
There is a thermal component of DM which is produced from the thermal plasma, and its number density follows equilibrium and then becomes frozen at around $T_{\rm fr} \simeq m_\chi/20$. However, the component are soon dominated by the non-thermal DM.

Even though $\Treh \ll T_{\rm fr}$  and thermally produced dark matters are already frozen, 
the non-thermal DMs can re-annihilate again into light particles, when their number density is large enough to satisfy
\dis{
n_\chi\VEV{\sigma_{\rm A} v} > H, \label{cond_reann}
}
where $\VEV{\sigma_{\rm A} v} \sim \sigma_A$ is the  total annihilation cross section of  non-thermal DM arising from the decay of $\phi$, which is relativistic with energy
$m_\phi/2$. \footnote{For complete calculations, we need to keep track of the momentum dependence of the DM distribution function~\cite{Kim:2016spf}. However, its effect is expected to be not so substantial to change our main results.  In our scenario, we restrict ourselves to $m_\chi\sim m_\phi$, in which case the  thermally averaged cross section of DM is similar to the non-thermall averaged one. }
The Hubble parameter $H$ is given by the total sum of the energy density in the Universe as
\dis{
 H^2 =\frac{1}{3\Mp^2} (\rho_\phi + \rho_r  +  \rho_\chi ),
}
where $\rho_\chi$ is the energy density of DM.

The Boltzmann equations which govern the evolution are written as
\begin{align}
\dot\rho_\phi + 3 H \rho_\phi & = - \Gamma_\phi \rho_\phi \, , \label{eq:rho_phi}
\\
\label{eq:rho_r}
\dot\rho_r + 4 H \rho_r & = (1-f_\chi) \Gamma_\phi \rho_\phi +2 \VEV{\sigma_A v} \bfrac{m_\phi}{2} n_\chi n_{\bar{\chi}}\, ,
\\
\dot{n}_\chi+3H n_\chi & =  f_\chi\Gamma_\phi \frac{\rho_\phi}{m_\phi} - \VEV{\sigma_A v}( n_\chi n_{\bar{\chi}} - n^{\rm eq}_\chi n^{\rm eq}_{\bar{\chi}} )\,  ,\label{eq:rho_chi}\\
\dot{n}_{\bar{\chi}}+3H n_{\bar{\chi}} & =  f_\chi\Gamma_\phi \frac{\rho_\phi}{m_\phi} - \VEV{\sigma_A v} ( n_\chi n_{\bar{\chi}} - n^{\rm eq}_\chi n^{\rm eq}_{\bar{\chi}} ) \,  ,\label{eq:rho_chib}
\end{align}
where $f_\chi$ is the branching ratio of $\phi$ decay to DM, e.g. $\phi \rightarrow \chi + \bar{\chi}$.

When the decay is the dominant source, the  approximate scaling solutions for $\phi$ and the radiation are given by
\dis{
\rho_\phi &= \rho_{\phi,i} \bfrac{a_i}{a}^3 e^{-\Gamma_\phi t},\\
\rho_r  &\simeq \frac{2}{5} \frac{(1-f_\chi)\Gamma_\phi}{H}\rho_\phi \propto a^{-3/2}.\label{apporx_sol}
}
For DMs, they follow the thermal equilibrium initially and soon freeze out settling into the quasi-stable state where the production from decay and the annihilation equals to each other. At this epoch, the scaling solutions are given by
\dis{
n_\chi &\simeq n_{\bar{\chi}} \simeq \bfrac{f_\chi\Gamma_\phi\rho_\phi}{ \VEV{\sigma_A v}  m_\phi}^{1/2} \propto a^{-3/2}. \label{sol_nchi}
}
After reheating, when there is no more production of non-thermal DM,  the DM annihilation is efficient and the final abundance is rearranged as~\cite{Nakamura:2007wr}
\dis{
Y_\chi \equiv \frac{n_\chi}{s} \simeq  \frac{H(\Treh)}{ \VEV{\sigma_{\rm A} v} s} \simeq \frac14 \bfrac{90}{\pi^2 g_*}^{1/2} {\frac{1}{\VEV{\sigma_{\rm A} v}\Mp \Treh }},
\label{YDM}
}
and the corresponding relic density of DM is
\dis{
\Omega_\chi h^2\simeq 0.14\bfrac{90}{\pi^2 g_*}^{1/2} \bfrac{m_\chi}{1\tev}\bfrac{10^{-8}\gev^{-2}}{\VEV{\sigma_{\rm A} v}}  \bfrac{20\gev}{\Treh}. \label{OmegaDM}
}

Since the non-thermal DMs are out of equilibrium, a baryon asymmetry can be generated during the re-annihilation of DMs. 
The CP asymmetry, $\epsilon$,  generated via the B number violating DM annihilations, can be parametrised as
\dis{
\epsilon = \frac{\sigma_\slashed{B}(\chi \chi \rightarrow \cdots) - \sigma_\slashed{B}(\bar{\chi}\bar{\chi} \rightarrow\cdots) }{\sigma_\slashed{B}(\chi\chi \rightarrow \cdots) + \sigma_\slashed{B}(\bar{\chi}\bar{\chi} \rightarrow\cdots) },
}
where $\sigma_\slashed{B}(\chi\chi \rightarrow \cdots)$  and $ \sigma_\slashed{B}(\bar{\chi}\bar{\chi} \rightarrow\cdots)$ are the B number violating annihilation cross sections of DM.

Then the Boltzmann equation for the baryon asymmetry $n_B$ is given by
\dis{
\dot{n}_B +3Hn_B = \epsilon \VEV{\sigma_{\slashed{B}} v}(n_\chi^2 - (n_\chi^{\rm eq})^2)-\VEV{\sigma_{\rm washout}v} n_B n_{\rm eq}  \, , \label{eq_B}
}
where $\sigma_{\slashed{B}}$ is the total B number violating annihilation cross section of DM,
\dis{
\sigma_{\slashed{B}} = \sigma_\slashed{B}(\chi\chi \rightarrow \cdots) + \sigma_\slashed{B}(\bar{\chi}\bar{\chi} \rightarrow\cdots),
}
which can be comparable to or smaller than the total annihilation of DM,
$ \VEV{\sigma_{\slashed{B}} v} \lesssim \VEV{\sigma_A v}$.
The last term in \eq{eq_B} is the washout effect, which  is expected to be the same order as the B number violating interaction, $\sigma_{\rm washout} \sim  \sigma_{\slashed{B}}$. 
For successful baryogenesis, however, the washout term must be suppressed.

\begin{figure*}[!t]
\begin{center}
\begin{tabular}{cc} 
 \includegraphics[width=0.45\textwidth]{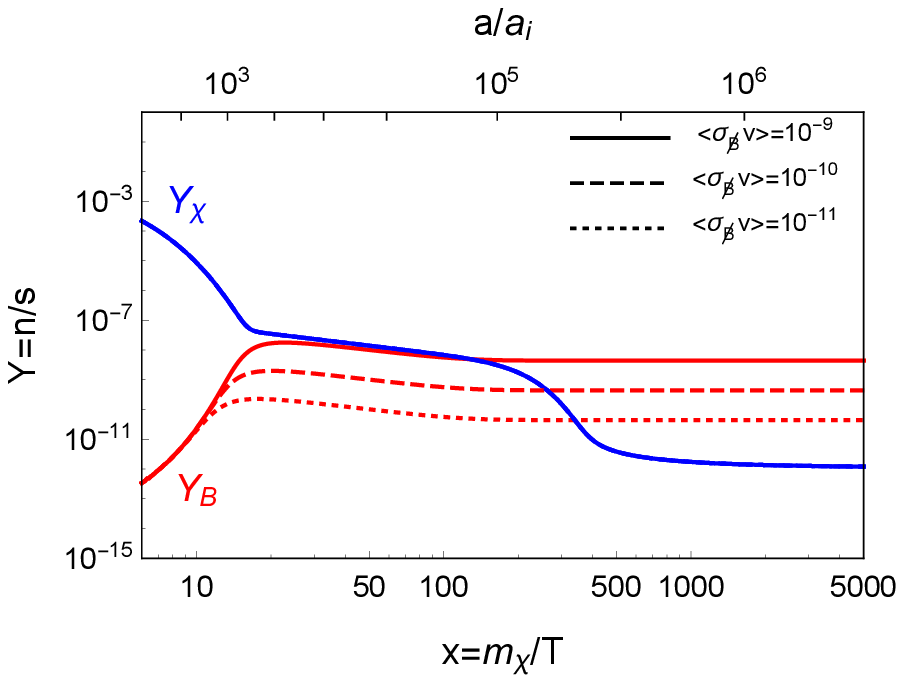}
&
 \includegraphics[width=0.45\textwidth]{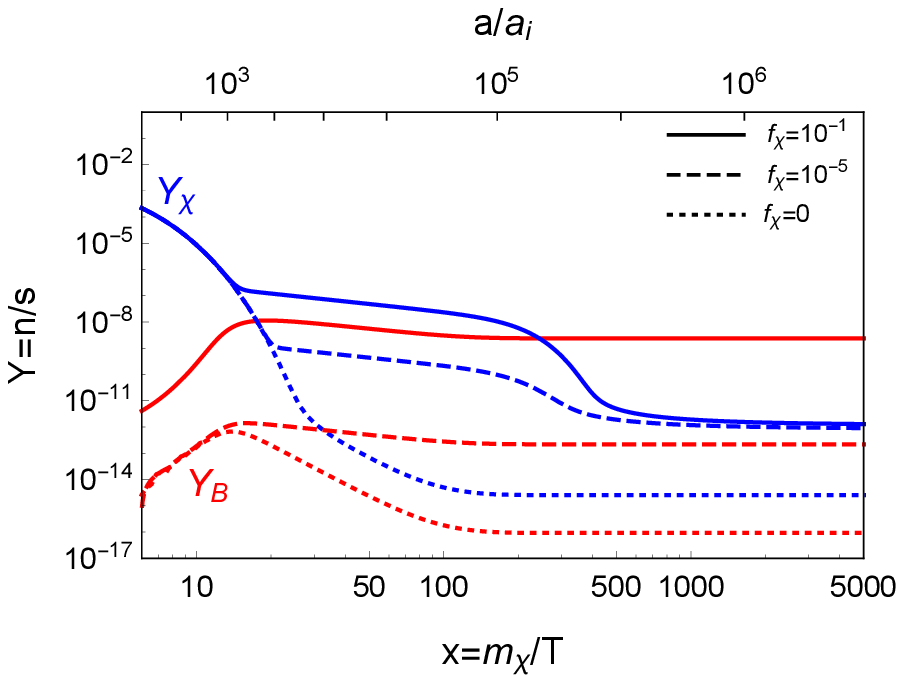}
  \end{tabular}
\end{center}
\caption{Plots of $Y_\chi$ and $Y_B$ for 
$\VEV{\sigma_{\slashed{B}} v} = 10^{-9}$ (solid), $10^{-10}$ (dashed), $10^{-11}$ (dotted)$ \gev^{-2}$  (left panel) and 
$f_\chi=10^{-1}$ (solid), $10^{-5}$ (dashed), $0$ (dotted) (right panel). Other parameters are fixed by the same values in Fig.~\ref{Fig:rho_Y}: $m_\phi =5\,{\rm TeV}$, $m_\chi=2\tev$, $\Treh = 20\,{\rm GeV}$, $\epsilon=0.001$, $m_\psi=3\tev$,   $ \VEV{\sigma_A v} =10^{-8}\,  {\rm GeV^{-2}}$,
 and  $f_\chi=0.01$ (Left) and  and $ \VEV{\sigma_{\slashed{B}} v} = 0.005 \times \VEV{\sigma_{A} v} $ (Right).
} 
	\label{Fig:Y}
\end{figure*}

The washout term can be suppressed in the case that 
\dis{
\frac{\VEV{\sigma_{\rm washout}v} n_B n_{\rm eq} }{\epsilon \VEV{\sigma_{\slashed{B}} v}n_\chi^2}  \ll 1.
}
This condition is satisfied when $n_B n_{\rm eq} \ll n^2_\chi$ and/or $\VEV{\sigma_{\rm washout}v} \ll \VEV{\sigma_{\slashed{B}} v}$.
The former can be achieved when one of the final particles, denoted by $\psi$ with mass $m_\psi$,  produced from the B number violating annihilation of DM is non-relativistic while keeping in thermal equilibrium and thus its number density, $n_{\rm eq}$, is exponentially suppressed as $e^{-m_\psi/T}$.

During the matter-dominated era by $\phi$, when ignored the washout effect, we can find the scaling solution for the baryon number density as
\dis{
n_B = \epsilon {\VEV{\sigma_{\slashed{B}} v}} n_\chi^2 \frac{2}{3H}=\frac{2 \epsilon f_\chi \Gamma_\phi\Mp}{\sqrt{3} m_\phi } \frac{\VEV{\sigma_{\slashed{B}} v}}{\VEV{\sigma_A v}} \rho_\phi^{1/2}\propto a^{-3/2}, \label{sol_nb}
}
where we have used \eq{sol_nchi} and $t=2/3H$ during matter-domination.

Now let us estimate the baryon asymmetry created in our scenario in the sudden decay approximation.
After reheating, the number density of the non-thermal DM is 
\dis{
n_\chi =f_\chi n_\phi =  f_\chi  \frac{\rho_\phi}{m_\phi} \, ,
}
with  the number density of $\phi$, $n_\phi=\rho_\phi/m_\phi$.
The DM re-annihilation can happen when the WIMP annihilation cross section satisfies the condition,
\dis{
f_\chi \VEV{\sigma_{\rm A} v}> \frac{m_\phi}{3\Mp \Treh^2}\bfrac{90}{\pi^2 g_*}^{1/2},
} from \eq{cond_reann}.
Therefore the baryon asymmetry is estimated as
\dis{
Y_B &\sim \epsilon \frac{\VEV{\sigma_{\slashed{B}} v}}{\VEV{\sigma_A v}} Y_\chi = \epsilon  f_\chi \frac{\VEV{\sigma_{\slashed{B}} v}}{\VEV{\sigma_A v}} \frac{n_\phi}{s}\\
& \sim 10^{-10} \bfrac{ \epsilon}{10^{-3}} \bfrac{ f_\chi }{10^{-2}} \bfrac{\VEV{\sigma_{\slashed{B}} v}/\VEV{\sigma_A v}}{10^{-2}} \bfrac{\Treh/m_\phi}{10^{-3}},
\label{Y_B}
}
 where $Y\equiv n/s$ with the entropy density $s=(2\pi^2/45g_{*S}) T^3$ and effective entropy degrees of freedom $g_{*S}$. In the second line in \eq{Y_B}, we have used
 \dis{
 \frac{n_\phi}{s} \simeq	 \frac34\frac{\Treh}{m_\phi}.
 }

\section{Numerical Results}
\label{results}
In the left panel of Fig.~\ref{Fig:rho_Y}, we show how  the background energy density of $\phi$ and the radiation evolve along with $x=m_\chi/T$ (or $a/a_i$)  for $\Treh = 20\, {\rm GeV}$, $m_\chi=2\tev$, and $m_\phi =5{\rm TeV}$. We can see that the reheating happens at around $x\simeq 100$ (or $a/a_i\simeq 10^5$) and the energy density of the radiation shows the scaling behavior decreasing proportional to $a^{-3/2}$  before reheating and  to $a^{-3}$  after reheating, as shown in the \eq{apporx_sol}.

In the right panel of Fig.~\ref{Fig:rho_Y},  we plot  $Y_\chi$  and $Y_B$  for  $\VEV{\sigma_A v} =10^{-7}$ (solid line), $10^{-8}$ (dashed line), $10^{-9}$ (dotted line) ${\rm GeV^{-2}}$.
We take  $\epsilon=0.001$ and $f_\chi=0.01$ as inputs, and the ratio, $\VEV{\sigma_{\slashed{B}} v}/\VEV{\sigma_A v}$, is fixed to be $5\times 10^{-3}$. 
Note that  DMs are in the thermal equilibrium initially and frozen  and soon become dominated by the non-thermal components produced from the decay of $\phi$ at around $a/a_i\simeq 10^3$ corresponding to $x=m_\chi/T \simeq 20$. During this period, the abundance $Y_\chi$ scales as
\dis{
Y_\chi = \frac{n_\chi + n_{\bar{\chi}}}{s}\sim \frac{a^{-3/2}}{T^3} \propto a^{-3/8},
}
where we have used \eq{sol_nchi} and adopted $T\propto a^{-3/2}$ during matter domination. This scaling can be seen in the right panel of Fig.~\ref{Fig:rho_Y}. The abundance of DM in the scaling regime depends on $\VEV{\sigma_A v}^{-1/2}$ as in \eq{sol_nchi}, however  after reheating, the DMs re-annihilate quickly and the final relic density  is inversely proportional to $\VEV{\sigma_A v}$ as in~\eq{YDM}. 

As mentioned above,  baryon asymmetry can be generated from the annihilatons of WIMP DM
when DM begins to deviate from the equilibrium and washout effect freezes out, that is called (thermal) WIMPy baryogenesis~\cite{Cui:2011ab,Kumar:2013uca,Racker:2014uga}. However, during matter-dominated era, the baryon asymmetry 
is soon dominated by that generated from the annihilation of non-thermal DM. During this period, the baryon asymmetry also shows the scaling behavior as in \eq{sol_nb}.  In the right panel of Fig.~\ref{Fig:rho_Y}, the abundance of baryon asymmetry, $Y_B$ appears independent of $\VEV{\sigma_A v}$ because the ratio $\VEV{\sigma_{\slashed{B}} v}/\VEV{\sigma_A v}$ is fixed in this figure, as can be seen in \eq{sol_nb}.
We can see from the figure that the required value for the baryon asymmetry $Y_B\sim 10^{-10}$ given in \eq{Ybaryon} can  be easily obtained. Here we used $m_\psi=3\tev$ to suppress the wash-out effect. The dependence of $Y_B $ on $m_\psi$ will be presented  in Fig.~\ref{Fig:psi}.

In Fig.~\ref{Fig:Y}, we show  the evolution of $Y_\chi$  and $Y_B$ 
 for different values of $\VEV{\sigma_{\slashed{B}} v}$ (left panel) and $f_\chi$ (right panel). We take $\VEV{\sigma_{\slashed{B}} v} = 10^{-9}$ (solid line), $ 10^{-10}$ (dashed line), $10^{-11}$ (dotted line)  $\gev^{-2}$   in the left panel and  $f_\chi=10^{-1}$ (solid line), $10^{-5}$ (dashed line), and $0$ (dotted line)   in the right panel for the same inputs as in Fig.~\ref{Fig:rho_Y},
 except that we chose $ \VEV{\sigma_A v} =10^{-8}\,  {\rm GeV^{-2}}$  
  and $ \VEV{\sigma_{\slashed{B}} v} = 0.005 \times \VEV{\sigma_{A} v}  $. 
One can easily see that the final $Y_B$ is proportional to not only $\VEV{\sigma_{\slashed{B}} v}$ in the left panel, but also $f_\chi$ in the right panel. 

In the limit of $f_\chi \rightarrow 0$, there is no DM production from the decay of heavier particle. However it does not simply lead to the result of thermal WIMP baryogenesis. Instead, the baryon asymmetry generated from the decay of $\phi$ during the WIMP freeze-out is diluted due to the entropy generation. This can be seen in the right figure of Fig.~\ref{Fig:Y}. In this case with $f_\chi= 0$, we find that $Y_B \simeq 10^{-16}$ and $Y_\chi \simeq 10^{-14}$, which is too small to explain baryogenesis and dark matter relic density simultaneously.

In Fig.~\ref{Fig:psi}, we show the effects of wash-out by changing the mass of a particle, $\psi$, which is produced in the B-violating DM annihilation.  Here we have used $m_\psi=3\tev$ (solid line), $1.5\tev $ (dashed line), $300\gev$ (dot-dashed line), 150$\gev$ (dotted line). 
We can clearly see from Fig.~\ref{Fig:psi}  that the wash-out is effective and thus  $Y_B$ is suppressed when $T\gtrsim m_\psi$, but at lower temperatures $T\lesssim m_\psi/25$ the wash-out is suppressed and  thus $Y_B$ can be sizable. 
For a given reheating temperature $\Treh=20\gev$, the final $Y_B$ is affected when $m_\psi \lesssim 500 \gev$, which is roughly $25 \Treh$.
Note that $Y_\chi$ is indendent of the washout effect.

\section{A model suppressing washout}
\label{model}
\begin{figure}[!t]
\begin{center}
\begin{tabular}{c} 
 \includegraphics[width=0.45\textwidth]{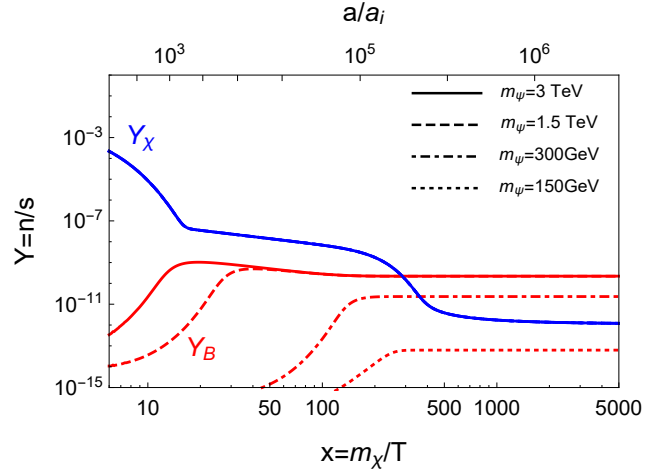}
 \end{tabular}
\end{center}
\caption{Plots of  $Y_\chi$  and $Y_B$  for different values of $m_\psi=3\tev$ (solid), $1.5\tev$ (dash-dotted), $300\gev$ (dashed),  $150\gev$ (dotted). 
Other model parameters are fixed with the same values in Fig.~\ref{Fig:Y} and $\VEV{\sigma_{\slashed{B}} v}= 5\times 10^{-3}\VEV{\sigma_{A} v} $.} 
	\label{Fig:psi}
\end{figure}

As a specific example for the suppression of washout effects, we adopt the model suggested in~\cite{Cui:2011ab,Bernal:2012gv,Bernal:2013bga}, where the DMs annihilate to quarks directly, and embed  the non-thermal WIMP baryogenesis in the model.
The model includes a vectorlike gauge singlet dark matter $X$ and $\bar{X}$, singlet pseudoscalars $S_\alpha$, and vectorlike exotic quark color triplets $\psi_i$ and $\bar{\psi}_i$, with an interaction
\dis{
\Delta \mathcal{L} = -\frac{i}{2}(\lambda_{X\alpha} X^2 + \lambda'_{X\alpha} \bar{X}^2  )S_\alpha
+i\lambda_{B\alpha}S_\alpha \bar{u}\psi  .
}
 The DM annihilations occurs through the intermediate $S$ states into $u$ and $\psi$, 
\dis{
XX\rightarrow S^* \rightarrow \bar{u} \psi, \quad {\rm and} \quad \bar{X}\bar{X} \rightarrow S^* \rightarrow  u \psi^\dagger.
}
At the same time, we assume that there are DM annihilations that do not violate baryon number.
A baryon asymmetry is generated in $u$ as well as in $\psi$. Since we consider $\psi$ is decoupled from the SM sector while in equilibrium with the hidden sector fields,  the baryon asymmetry in the SM field is not eliminated.
We assume that the reheating temperature is lower than the mass scale of $\psi$, $\Treh \ll m_\psi$, so that
the washout effect is suppressed thanks to the exponential suppression of  the number density of $\psi$ as shown in Fig.~\ref{Fig:psi}.

The CP asymmetry is given  by~\cite{Cui:2011ab}
\dis{
\epsilon \simeq -\frac{1}{6\pi}\frac{Im(\lambda^2_{B1} \lambda^{*2}_{B2})}{|\lambda_{B1}|^2} \,\label{epsilon_Cui}
}
where we assumed that $m_\phi \gg m_\psi$.
For  the couplings $\lambda_X=1$ and  $ \lambda_{B1}=\lambda_{B2} = 0.1$, the CP asymmetry given by
\eq{epsilon_Cui}  is around $\epsilon \sim 2.5\times 10^{-4}$ in the case of the maximal CP violating phase.
The B violating DM annihilation cross section is estimated as
$\sigma_\slashed{B} \sim (\lambda_X \lambda_B)^2 \frac{1}{m_S^2}$.

\begin{figure}[!t]
\begin{center}
\begin{tabular}{c} 
 \includegraphics[width=0.45\textwidth]{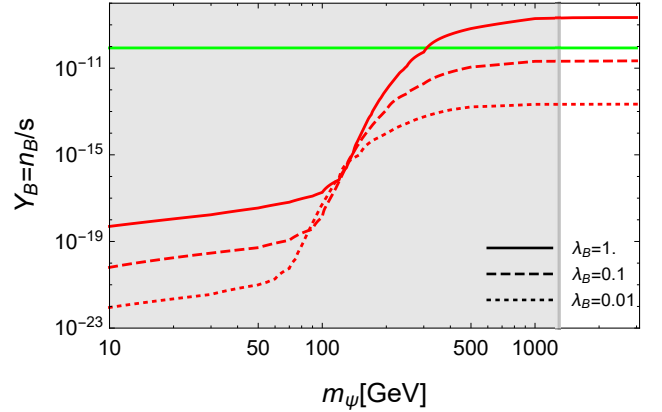}
  \end{tabular}
\end{center}
\caption{Plot of $Y_{B}$ versus $m_\psi$ for different model parameters in the Section \ref{model}:  $\lambda_X=1$ and  $\lambda_B=$1 (solid), 0.1 (dashed), 0.01 (dotted) with $m_S=10\tev$, $m_\chi=2\tev$, and $\Treh=20\gev$. The green horizontal line is the measured baryon abundance in \eq{Ybaryon} and shaded region is disfavored by the collider experiment. }
	\label{Fig:model}
\end{figure}

In Fig.~\ref{Fig:model}, we show the effects of the wash-out depending on  $m_\psi$ for different model parameters, $\lambda_B \equiv \lambda_{B1}=\lambda_{B2}= 1$ (solid line), 0.1 (dashed line), 0.01 (dotted line) and $\lambda_X=1$ for $\Treh=20\gev$ and $m_\chi=2\tev$.
As can be seen here,  the washout of $Y_B$ is sizable for  $m_\psi \lesssim \Treh$, whereas that  is ineffective for $m_\psi \gtrsim 500\gev $ which corresponds roughly to $25 \Treh$. The horizontal line represents the required baryon asymmetry in \eq{Ybaryon}.

In this model, since the particle $\psi$ has color charges, $m_\psi$  is strongly constrained by the collider searches. From the current LHC gluino search, it is inferred that the lower bounds on $m_\psi$ are $ 1.3 \tev - 1.6\tev$ depending on neutralino mass ~\cite{Khachatryan:2016uwr}. The shaded region in Fig.~\ref{Fig:model} is disfavored by those bounds. For sufficient suppression of wash-out, we require that the reheating temperature is smaller than $m_\psi/25$. At the same time, if we assume that the reheating temperature is smaller than the thermal freeze-out temperature of DM,
then we need $\Treh < m_\chi/25$.

The  relic density of dark matter is determined as in~\eq{OmegaDM} which depends on the total annihilation cross section of dark matter as well as the reheating temperature.
For $\Treh = 20\gev$, and $m_\chi=2\tev$,  the right value of the DM relic density can be obtained with $\VEV{\sigma_{\rm A} v} \simeq  5 \times 10^{-8}  \gev^{-2}$.
This gives a testable window  in the indirect detection of dark matter depending on the annihilation modes. If the dominant annihilation modes of dark matter is $b\bar{b}$ then it might be marginally possible to detect signals in the gamma-ray detection ~\cite{Ackermann:2015zua}. On the other hand, if the dominant annihilation is into the hidden sector, then it might be difficult to see signals through the indirect searches.

\section{Conclusion}
\label{Con}
In this work, we have proposed a WIMP baryogenesis that can be reconciled with low reheating temperature.
In this scenario, DM is non-thermally produced during a reheating period created by the decay of long-lived heavy particle, and subsequently re-annihilate to lighter particles even after the thermal freeze-out. The re-annihilation of DM provides the observed baryon asymmetry  as well as the correct relic density of DM.  
We have investigated how wahout effects can affect the generation of the baryon asymmetry and studied a model suppressing them by introducing a heavy particle $\psi$, which is decoupled from the SM fields
when it is produced.
From the analysis, we have found that DM can be heavy enough and its annihilation cross section  can  also be larger than that adopted in the usual thermal WIMP baryogenesis.

\section*{Acknowledgment}
K.-Y.C. and J.Kim was supported by the National Research Foundation of Korea(NRF) grant funded by the Korea government(MEST) (NRF-2016R1A2B4012302). S.K.Kang and J.Kim was  supported by the National Research Foundation of Korea(NRF) grant (2009-0083526,  2017K1A3A7A09016430) and S.K.Kang was supported by  the NRF grant(2017R1A2B4006338).
J. Kim was supported by the National Research Foundation of Korea (NRF) grant funded by the Korea government (MEST) (NRF-2015R1D1A1A01061507).

%


\def\prp#1#2#3{Phys.\ Rep.\ {\bf #1} #2 (#3)}
\def\rmp#1#2#3{Rev. Mod. Phys.\ {\bf #1}  #2 (#3)}
\def\anrnp#1#2#3{Annu. Rev. Nucl. Part. Sci.\ {\bf #1} #2 (#3)}
\def\npb#1#2#3{Nucl.\ Phys.\ {\bf B#1}  #2 (#3)}
\def\plb#1#2#3{Phys.\ Lett.\ {\bf B#1}  #2 (#3)}
\def\prd#1#2#3{Phys.\ Rev.\ {\bf D#1}, #2  (#3)}
\def\prl#1#2#3{Phys.\ Rev.\ Lett.\ {\bf #1}  #2 (#3)}
\def\jhep#1#2#3{JHEP\ {\bf #1}  #2 (#3)}
\def\jcap#1#2#3{JCAP\ {\bf #1}  #2 (#3)}
\def\zp#1#2#3{Z.\ Phys.\ {\bf #1}  #2 (#3)}
\def\epjc#1#2#3{Euro. Phys. J.\ {\bf #1}  #2 (#3)}
\def\ijmp#1#2#3{Int.\ J.\ Mod.\ Phys.\ {\bf #1}  #2 (#3)}
\def\mpl#1#2#3{Mod.\ Phys.\ Lett.\ {\bf #1}  #2 (#3)}
\def\apj#1#2#3{Astrophys.\ J.\ {\bf #1}  #2 (#3)}
\def\nat#1#2#3{Nature\ {\bf #1}  #2 (#3)}
\def\sjnp#1#2#3{Sov.\ J.\ Nucl.\ Phys.\ {\bf #1}  #2 (#3)}
\def\apj#1#2#3{Astrophys.\ J.\ {\bf #1}  #2 (#3)}
\def\ijmp#1#2#3{Int.\ J.\ Mod.\ Phys.\ {\bf #1}  #2 (#3)}
\def\apph#1#2#3{Astropart.\ Phys.\ {\bf B#1}, #2 (#3) }
\def\mnras#1#2#3{Mon.\ Not.\ R.\ Astron.\ Soc.\ {\bf #1}  #2 (#3)}
\def\nat#1#2#3{Nature (London)\ {\bf #1}  #2 (#3)}
\def\apjs#1#2#3{Astrophys.\ J.\ Supp.\ {\bf #1}  #2 (#3)}
\def\aipcp#1#2#3{AIP Conf.\ Proc.\ {\bf #1}  #2 (#3)}




\end{document}